# Nondegenerate and almost hexagonal skyrmion lattices

J. C. Martinez and M. B. A. Jalil

*Computational Nanoelectronics and Nano-device Laboratory*
*National University of Singapore, 4 Engineering Drive 3, Singapore 117576.*

We obtain the lowest energy solutions for the skymion field equations and their corresponding vortex structures. Two nondegenerate solutions emerge with their vortex swirls in opposite directions. The solutions are associated with an extremum property, which favors an array of almost hexagonal shape. We predict that a *regular* hexagonal lattice must have a mix of skyrmions of both swirls. Although our solutions could not keep the norm of the magnetization constant at unity, their greatest deviation from unity occurred in regions where the spins are far from planar; we show how to improve this situation.

In chiral ferromagnets, i.e., metallic magnets lacking inversion symmetry, the characteristic ferromagnetic exchange interaction $J$ and the weaker Dzyaloshinkii-Moriya (DM) coupling $D$ conspire to create a helical spin ground state with pitch-vector length $2\kappa \equiv D/J$ [1]. In the environment of an external magnetic field **B** perpendicular to the thin-film plane, the spiral spin configuration for **B** = 0 is replaced by a hexagonal packing of skyrmions and above a critical external field $\mathbf{B}_c$ the skyrmion crystal turns into a fully polarized ferromagnet [2]. Both transitions are of first order. At a region intermediate between these phases an Abrikosov-like vortex array, as in type-II superconductors, becomes manifest. This finding appears to confirm a hypothesis put forward by Rößler et al [3]. For recent reviews, see [4- 6]. Besides these long-awaited findings, interest in the skyrmion crystal has been fueled by the recent discoveries of induced skyrmion motion at remarkably low current densities [7, 8] and control through magnons [9] as well as the lively debate on the race track memory [10].

In this letter we show the existence of two distinct and nondegenerate skyrmion lattices depending on the sense of swirl. The energy difference between the configurations per skyrmion pair is estimated at 2 meV. Comparison of this value with the typical energy gap of ~1 eV in ferromagnets could explain the fact of low current densities inducing skyrmion motion mentioned above. We also find that the skyrmion lattice is almost but *not* quite a regular hexagon, as is generally assumed. Both these results are found to be directly related to the DM interaction, assigning to it a crucial role in chiral ferromagnets: although the DM interaction is small compared with the exchange interaction it plays an important and delicate function of oversight in the distribution of energy within the crystal. The two distinct lattice configurations exist because of the DM interaction.

A standard starting point is the Belavin-Polyakov O(3) nonlinear sigma model whose energy in the noninteracting case is given by $E = \frac{1}{2}\int d^2x(\partial_i n^a)^2$, where the 'order parameter' $n^a = (\sin\theta\cos\phi, \sin\theta\sin\phi, \cos\theta)$ is a three-dimensional spin (unit) vector; $i = 1,2$; $a = 1,2,3$ [11]. Since the configuration space is multiply-connected, as exemplified by the homotopy $\pi_3(S^2) = Z$, it is convenient to employ a $CP^1$ description through a two-component spinor $\mathbf{z} = \begin{pmatrix} z_1 \\ z_2 \end{pmatrix}$, of unit normalization and the identification $\mathbf{n} = \mathbf{z}^\dagger \boldsymbol{\sigma} \mathbf{z}$, where $\boldsymbol{\sigma}$ is the Pauli matrix [12]. In the interacting case with an external magnetic field $\mathbf{B}$ pointing upward, the Belavin-Polyakov energy is replaced by the Ginzburg-Landau free energy [13]

$$E = \int d^2r\{2J(D_\mu \mathbf{z})^\dagger(D_\mu \mathbf{z}) - \mathbf{B}\cdot \mathbf{z}^\dagger \boldsymbol{\sigma} \mathbf{z}\} \qquad (1)$$

in which the spatial integration is over the two-dimensional plane ($\mu = x, y$), the covariant derivative $D_\mu = \partial_\mu - iA_\mu + i\kappa\sigma_\mu$ is a 2×2 matrix, the spin vector is related to the complex field via $\mathbf{n} = \mathbf{z}^\dagger \boldsymbol{\sigma} \mathbf{z}$, and $A_\mu = -i\mathbf{z}^\dagger(\partial_\mu \mathbf{z})$ is an associated vector potential, non-locally dependent on $\mathbf{n}$. The saddle-point equation $\frac{\delta E}{\delta \mathbf{z}^\dagger} = 0$ yields the equation

$$2J(\boldsymbol{\nabla} - i\mathbf{A} + i\kappa\boldsymbol{\sigma})^2 \mathbf{z} + 2iD(\mathbf{n}\cdot\boldsymbol{\nabla})\mathbf{z} + (\mathbf{B}\cdot\boldsymbol{\sigma})\mathbf{z} = 0 \qquad (2)$$

A uniform field $H\,\hat{\mathbf{k}}$ induced by $\mathbf{A}$ is also introduced, which in Landau gauge is $A_x = 0$, $A_y = Hx$. Much of what follows is an attempt to solve Eq. (2). Han et al. [13] carried out an extensive analysis of Eq. (2) and obtained vortex solutions of the Abrikosov type but glossed over two important issues: their analysis (a) left out the second term of Eq. (2) and (b) ignored the normalization of $\mathbf{z}$. In taking up these issues we show how a complete solution of Eq. (2) can be obtained although we only obtain an approximate one; in fact we find a second solution not previously suspected. These solutions are shown to be associated with an energy extremum. Although we are unable to keep the normalization of **z** fixed at unity, we find that its greatest deviation from unity occurs where the magnetic moments are far from planar to the film. We also find that adding a correction term to **z** can improve the normalization and this correction is precisely due to the middle term of Eq. (2), which was neglected by Han et al.

Based on the general result $\mathbf{z} = \left(\cos\frac{\theta(\rho)}{2}e^{i\chi} \quad \sin\frac{\theta(\rho)}{2}e^{i(\phi+\chi)}\right)^T$, $0 \leq \theta \leq \pi$, $0 \leq \phi, \chi \leq 2\pi$, $\rho$ = radial distance [12], a simple form of the vector potential associated with **z** is $\boldsymbol{A} = -\frac{1+\cos\theta(\rho)}{2\rho}\hat{\mathbf{e}}_\phi$, which we recognize as the vector potential of a magnetic monopole of strength $\frac{-1}{2}$ [14] and which reduces on the *xy*-plane to $\boldsymbol{A} = -\frac{1}{2}\nabla\phi$. From $\boldsymbol{A} = -\frac{1+\cos\theta}{2\rho}\hat{\mathbf{e}}_\phi$, the associated magnetic field is $\mathbf{B} = \nabla \times \boldsymbol{A} = \frac{\sin\theta(\rho)}{2\rho}\frac{d\theta}{d\rho}\hat{\mathbf{z}}$. Then the flux of **B** over the plane is found to be $2\pi$. We interpret this result as the quantization of magnetic flux for a single skyrmion. (But note that there is no flux quantum in terms of $\hbar, c, e$.)

It is convenient to introduce $\boldsymbol{\Psi} = e^{i\frac{1}{2}\phi}\mathbf{z}$ and recast Eq. (2) (including the uniform field $H\,\hat{\mathbf{k}}$) as

$$\begin{pmatrix} \partial_x & i\kappa \\ i\kappa & \partial_x \end{pmatrix}^2 \boldsymbol{\Psi} + \begin{pmatrix} \partial_y - iHx & \kappa \\ -\kappa & \partial_y - iHx \end{pmatrix}^2 \boldsymbol{\Psi} + 2b\kappa^2 \begin{pmatrix} 1 & 0 \\ 0 & -1 \end{pmatrix}\boldsymbol{\Psi} + \frac{\ell}{2J}\boldsymbol{\Psi} + 2i\kappa\mathbf{n}\cdot(i\boldsymbol{A}+\nabla)\boldsymbol{\Psi} = 0 \quad (3)$$

in which $\kappa = \frac{D}{2J}$; $\frac{B}{2J} = 2b\kappa^2$, $\Lambda = \frac{\ell}{2J}$ and $b = BJ/D^2$ is dimensionless. The constant $\ell$ is a Lagrange parameter inserted above to ensure proper normalization of **z**. Since $\mathbf{n} = \mathbf{z}^\dagger\boldsymbol{\sigma}\mathbf{z}$ is the physical variable, the half-angle in $\boldsymbol{\Psi}$ does not cause alarm. For now, we ignore the last term of Eq. (3) and return to it later. At this point the DM interaction is present through the parameter $\kappa$ but its violation of inversion symmetry is ignored. Let $\Psi = e^{iky}\begin{pmatrix}F(x)\\iG(x)\end{pmatrix}$ and define $\sqrt{H}X = Hx - k$. Then we have

$$\begin{pmatrix}\partial_X^2 - X^2 - 2(1-b)\bar{\kappa}^2 + \bar{\lambda} & 2i\bar{\kappa}(\partial_X - X)\\ 2i\bar{\kappa}(\partial_X + X) & \partial_x^2 - X^2 - 2(1+b)\bar{\kappa}^2 + \bar{\lambda}\end{pmatrix}\begin{pmatrix}F(x)\\iG(x)\end{pmatrix} = 0 \tag{4}$$

where $\bar{\kappa} = \frac{\kappa}{\sqrt{H}}, \bar{\lambda} = \frac{\Lambda}{H}$ are dimensionless. We choose the length scale $l_H = \frac{1}{\sqrt{H}}$. Equation (4) yields the solutions: $\Psi = \begin{pmatrix}f_n\phi_{n+1}(X_m)\\-ig_n\phi_n(X_m)\end{pmatrix}e^{2\pi i m y/l_y}$, in which $X_m \equiv \frac{x - m l_x}{l_H}$ is dimensionless, $l_x = \frac{k}{H}, l_y = 2\pi/k$; and $m$ is any positive or negative integer. Here $\phi_n$ are the normalized wave functions of the one-dimensional harmonic oscillator. The length scales $l_x$ and $l_y$ are connected via $l_x l_y = 2\pi l_H^2$ and the coefficients $f_n$ and $g_n$ are related through $\frac{f_n}{g_n} = \frac{2\bar{\kappa}\sqrt{2}\sqrt{2n+1}}{-1+2\bar{\kappa}^2 b \pm \sqrt{8\bar{\kappa}^2(2n+1)+(2b\bar{\kappa}^2-1)^2}}$. This last result implies *two* solutions for every $n$ value and fixed external field. Then the lowest energy ($n = 0$) solution of Eq. (4) can be given as

$$\Psi = N\sum_{m=-\infty}^{\infty}C_m\begin{pmatrix}f_0\sqrt{2}(\frac{x-ml_x}{l_H})\\-ig_0\end{pmatrix}e^{2\pi i m y/l_y}e^{-(\frac{x-ml_x}{l_H})^2/2}, \tag{5}$$

in which the $C_m$ are arbitrary constants and $N$ is an overall normalization constant. The parameter $\bar{\lambda}_n$, found to be $\bar{\lambda}_n = 2(1 + 2n + \bar{\kappa}^2) \pm \sqrt{1 + 4\bar{\kappa}^2(2 + 4n + b(-1 + b\bar{\kappa}^2))}$ depends only on the index $n$. In the following we will need only the $n = 0$ results.

In the spirit of Abrikosov's method [15] periodicity in the $x$-direction is achieved by imposing the recursion relation $C_{n+\tilde{N}} = C_n$ for all $n$. For a triangular array of skyrmions we choose $\tilde{N} = 1$, $C_{2n} = 1$ and $C_{2n+1} = i$. For this case and with the aid of the Poisson summation formula [16] we can sum up the series (5) to obtain (the index 0 is a reminder that we have neglected the last term of Eq. (3))

$$\Psi_0 = C \begin{pmatrix} -f_0\sqrt{2}l_H \frac{d}{dx}\left\{ e^{-\frac{x^2}{2l_H^2}} \vartheta_3[\frac{2\pi}{l_y}i(x+iy), e^{-\frac{2l_x^2}{l_H^2}}] + ie^{-\frac{(x-l_x)^2}{2l_H^2}+\frac{2\pi iy}{l_y}} \vartheta_3[\frac{2\pi}{l_y}i(x-l_x+iy), e^{-\frac{2l_x^2}{l_H^2}}] \right\} \\ -ig_0\{ e^{-\frac{x^2}{2l_H^2}} \vartheta_3[\frac{2\pi}{l_y}i(x+iy), e^{-\frac{2l_x^2}{l_H^2}}] + ie^{-\frac{(x-l_x)^2}{2l_H^2}+\frac{2\pi iy}{l_y}} \vartheta_3[\frac{2\pi}{l_y}i(x-l_x+iy), e^{-\frac{2l_x^2}{l_H^2}}] \} \end{pmatrix} \quad (6)$$

in which $\vartheta_3(u,q)$ in an elliptic theta function and $C$ a constant.

Let us now return to the last term of Eq. (3), which we omitted. Suppose we write the complete solution of Eq. (3) as $\Psi = \sum_{n=-\infty}^{\infty} e^{ikny}\left(C_n\psi_n(x) + \psi_n^{(1)}(x)\right)$ where the first or zeroth-order term is just $\Psi_0$ above and $\psi_n^{(1)}$ represent the correction. Substituting into Eq. (3), we obtain an equation for the correction in the form

$$\begin{pmatrix} \partial_x & i\kappa \\ i\kappa & \partial_x \end{pmatrix}^2 \psi_n^{(1)} + \begin{pmatrix} \partial_y - iHx & \kappa \\ -\kappa & \partial_y - iHx \end{pmatrix}^2 \psi_n^{(1)} + 2b\kappa^2 \begin{pmatrix} 1 & 0 \\ 0 & -1 \end{pmatrix} \psi_n^{(1)} + \frac{\ell}{2J}\psi_n^{(1)} =$$

$$-2i\kappa \Psi_0^\dagger \sigma \Psi_0 \cdot (i\mathbf{A} + \nabla)\Psi_0 \quad (7)$$

This is an inhomogeneous equation for $\psi_n^{(1)}$. A solution exists provided the right-hand side of Eq. (7) is orthogonal to the solution of the homogenous equation (i.e. the left-hand side). Now the solution of the homogeneous equation is precisely just $\Psi_0$ so the condition in question becomes $\int d^2r\, \Psi_{0p}^\dagger (\Psi_0^\dagger \sigma \Psi_0) \cdot (i\mathbf{A} + \nabla)\Psi_0 = 0$, where $\Psi_{0p}$ is the $p^{\text{th}}$ term of the sum (5). This can be recast as a derivative, $\frac{\partial}{\partial C_p^*} \int d^2r\, \Psi_0^\dagger \sigma \Psi_0 \cdot \Psi_0^\dagger (i\mathbf{A} + \nabla)\Psi_0 = 0$, or, equivalently, as

$$-2i\kappa \int d^2r\, \Psi_0^\dagger \sigma \Psi_0 \cdot \Psi_0^\dagger (i\mathbf{A} + \nabla)\Psi_0 \text{ is an extremum.} \quad (8)$$

If we fall back on the original variable **z** we may identify this as $2\kappa \int d^2r\, \mathbf{n} \cdot \mathbf{A} \equiv 2\kappa G$ (in this formula **A** includes both vector potential $-\frac{1}{2}\nabla\phi$ and $\Psi_0^\dagger(i\nabla)\Psi_0$). From Eqs. (1) and (2) we can show that the energy is proportional to $\int d^2R\{2\bar\kappa(-\mathbf{n}\cdot\mathbf{A}) + \bar\lambda\}$. Now since $\bar\lambda$ depends only on $n$, we may exclude it and conclude that Eq. (8) is a criterion for an energy extremum. By analogy with the formula $U = \frac{1}{2}\int \mathbf{j}\cdot\mathbf{A}d^2r$, for the energy of steady currents [17], $G$ leads us to think of **n** as a current. Tracing back the origin of $G$, we discover that it comes directly from the DM interaction and the extremum property (8) is an indication of its role as the 'energy manager' of the crystal.

Turning now to numerical results, we follow Tokura and Nagaosa [4] and choose $Fe_{1-x}Co_xSi$ ($x = 0.1$), which has a transition temperature of 11 K and a helical period of $\lambda$=43 nm. Setting $2\kappa = \frac{D}{J} = \frac{2\pi}{\lambda} \approx \frac{1}{10a}$ we estimate a lattice constant of $a$ = 6.8 Å. Also based on the definition $\frac{B}{2J} \equiv 2b\kappa^2$, and estimates of Zang *et al.*[18] we set $b$ = 3/2. We have also put $\bar\kappa \cong \sqrt{1/2}$ and fixed the length scale at $l_H = \frac{1}{\sqrt{H}} \approx 9.5$ nm. From these values the ratio $\frac{f_0}{g_0} = 0.78, -1.28$ for $n$ = 0 is obtained, and we will refer to these as the *first and second* solutions, respectively. (Recall that for every $n$ there are two independent solutions.) Given the relation $l_x l_y = 2\pi l_H^2$ and the requirement $l_y = \frac{\sqrt{3}}{2}l_x$ for a triangular array of skyrmions we find $l_x$ =2.693, $l_y$ = 2.333 for a regular hexagonal skyrmion lattice.

The extremum property of $G \equiv \int d^2r\, \mathbf{n}\cdot\mathbf{A}$ discussed above now also be calculated numerically and we find a minimum for the first solution at the ratio (or *shape parameter*) $r \equiv l_y/l_x \cong 0.91$, which is close to the ratio for a hexagonal lattice of $\frac{\sqrt{3}}{2} \cong 0.87$. See Fig. 1, which displays results for both the first and second solutions. We observe that the curves cross the horizontal axis at the same points, alternate in attaining their maximum and minimum values: in fact, when the first is a minimum the second is a maximum and *vice versa*. This suggests that the skyrmion lattice may undergo switching from one solution to the other. Based on an estimate of the exchange constant for $Fe_{1-x}Co_xSi$ ($x = 0.1$) of $J \approx 0.6$ meV [19] the energy difference at $r = 0.91$ is $4\bar{\kappa}J\Delta G \approx 2$ meV. Note that other extrema are given in Fig. 1 but here we will focus only on the second.

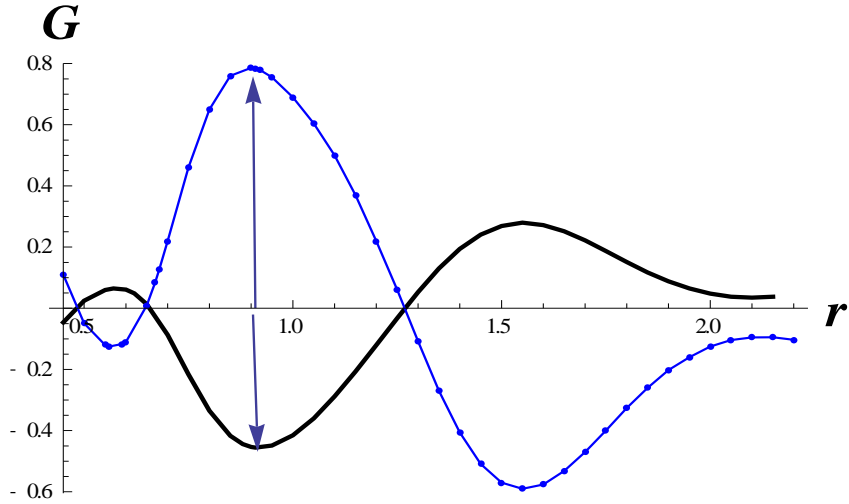

**Fig. 1** Plots of $G$ versus $r$, the ratio of $l_y/l_x$. The full thick curve corresponds to the first solution while the dot-connected curve to the second. The first minimum of the first solution occurs close to the ratio $\sqrt{3}/2 \cong 0.87$ for a triangular array. The arrow shown corresponds to an energy gap of $\Delta G \approx 1.25$ between the two solutions at $r = 0.91$.

Figure 2 displays the vortex structures for the first (second) solutions for $r \approx 0.91$, respectively. This value of $r$ corresponds to the second pair of extrema shown in Fig. 1. Note the opposite sense of the swirls of the two configurations. Thinking of **n** as a current as suggested above, we can understand the energy difference between the two configurations in the presence of a *fixed* magnetic field. Observe also the resulting *almost* hexagonal lattice. In Fig. 3 we give similar results for the $r = \sqrt{3}/2$ case, which is the hexagonal case. However this value of $r$ does *not* correspond to an energy minimum in Fig. 1. If, in a *regular* hexagonal lattice, we had suitable numbers of skyrmions of *both* solutions, with their corresponding swirls, it would then be possible to satisfy the energy extremum condition. This would not pose problems of continuity since at the edge of *any* skyrmion all the spins point upward. Thus we predict that a regular hexagonal lattice must have a mix of skyrmions of both swirls.

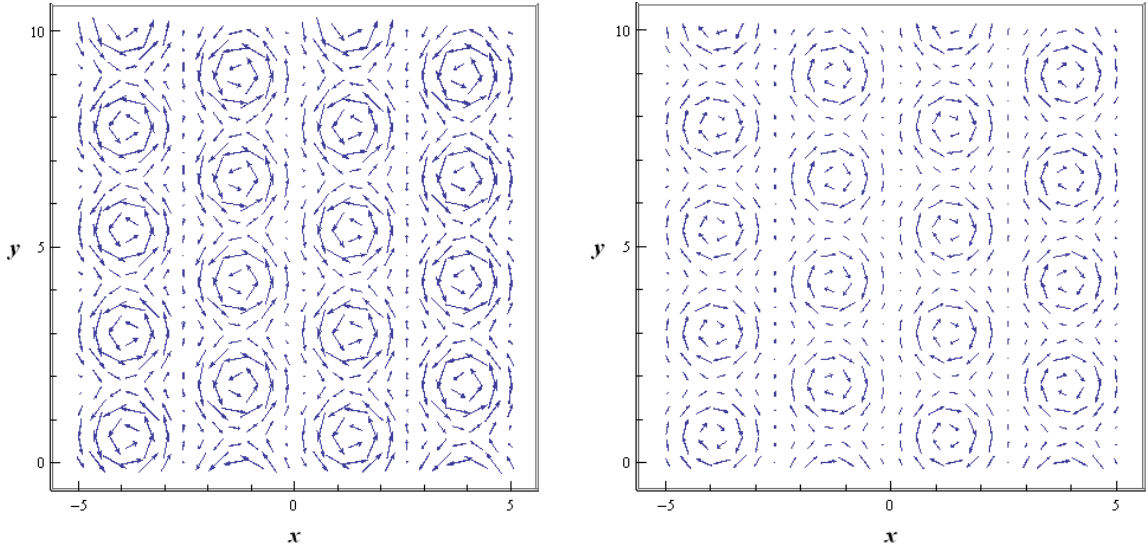

**Fig 2** Vortex structure corresponding to the first (left) and second (right) solutions for $r = 0.91$. The left graph is for $f_0 = 0.78$, while the graph on the right is for $f_0 = -1.28$. Length scale is $l_H = \frac{1}{\sqrt{H}} \approx 9.5$ nm.

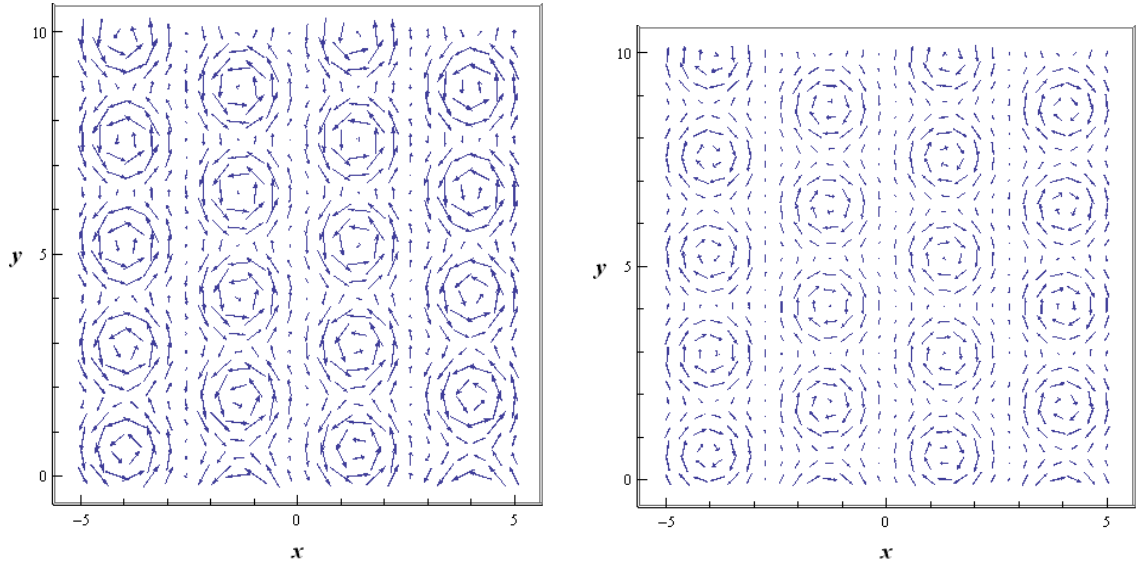

**Fig. 3** The same as Fig. 2 but for $r = \sqrt{3}/2$, the hexagonal case.

We show in Fig. 4 the configuration for a single skyrmion and the corresponding graph for the absolute value of the normalization |**n**| for the left plot of Fig. 2. Ideally this normalization is unity throughout space. We assume that |**n**| is close to unity in the region between the skyrmion's central core and its outer perimeter. We know that **n** is normal to the *xy* plane in the core and the perimeter. Examination of the right-hand plot shows that the greatest deviations from unity occur where the spins are pointing close to the normal (in the same direction or opposite).

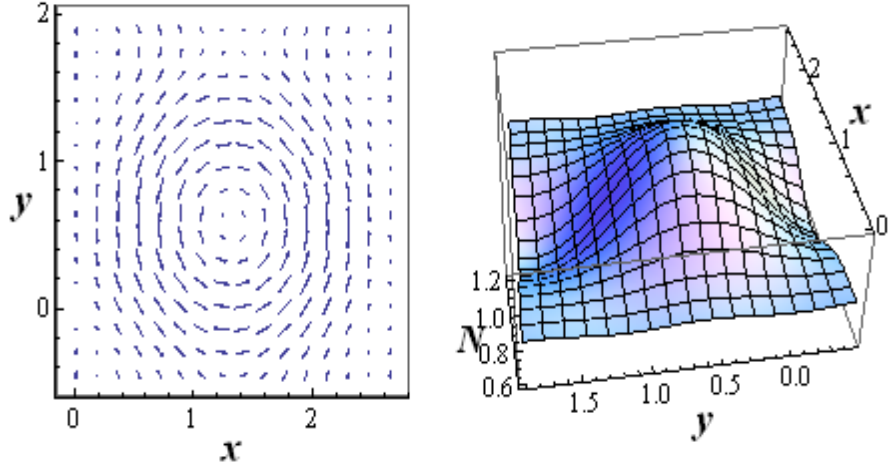

**Fig. 4** Vortex structure and absolute value *N* of the normal for a *single* skyrmion corresponding to the left-hand solution $\Psi_0$ of Fig. 2.

To go beyond these results we attempt to solve Eq. (7) approximately by writing it as

$$\begin{pmatrix} \partial_X^2 - X^2 - 2(1-b)\bar{\kappa}^2 + \bar{\lambda} & -2ia^\dagger \\ 2ia & \partial_x^2 - X^2 - 2(1+b)\bar{\kappa}^2 + \bar{\lambda} \end{pmatrix} \begin{pmatrix} F^{(1)}(x) \\ iG^{(1)}(x) \end{pmatrix} = -2i\bar{\kappa}\Psi_0^\dagger \sigma \Psi_0 \cdot$$

$(i\,\mathbf{A} + \nabla_R)\Psi_0$  (9)

In effect we are taking just *n* = 0 for $\psi_n^{(1)}$. Neglecting the gauge potential **A** (we have verified that its contribution is small) as well as all derivatives on the left-hand side and replacing $\Psi_0^\dagger \boldsymbol{\sigma} \Psi_0$ by its x-component, $n_x$, we simplify the above to $\begin{pmatrix} \bar{\lambda} & 0 \\ 0 & \bar{\lambda} \end{pmatrix} \begin{pmatrix} F^{(1)}(x) \\ iG^{(1)}(x) \end{pmatrix} \approx -i\sqrt{2} n_x \cdot \frac{d}{dx}\Psi_0$, in which $\bar{\lambda}_0 = 5.06$ for the first solution. This gives an expression for the correction term. This can now be used to redraw the vortex structure for the first solution, which is shown in Fig. 5. The right-most plot shows that the normalization for this case is better than that for the uncorrected case (see Fig. 4).

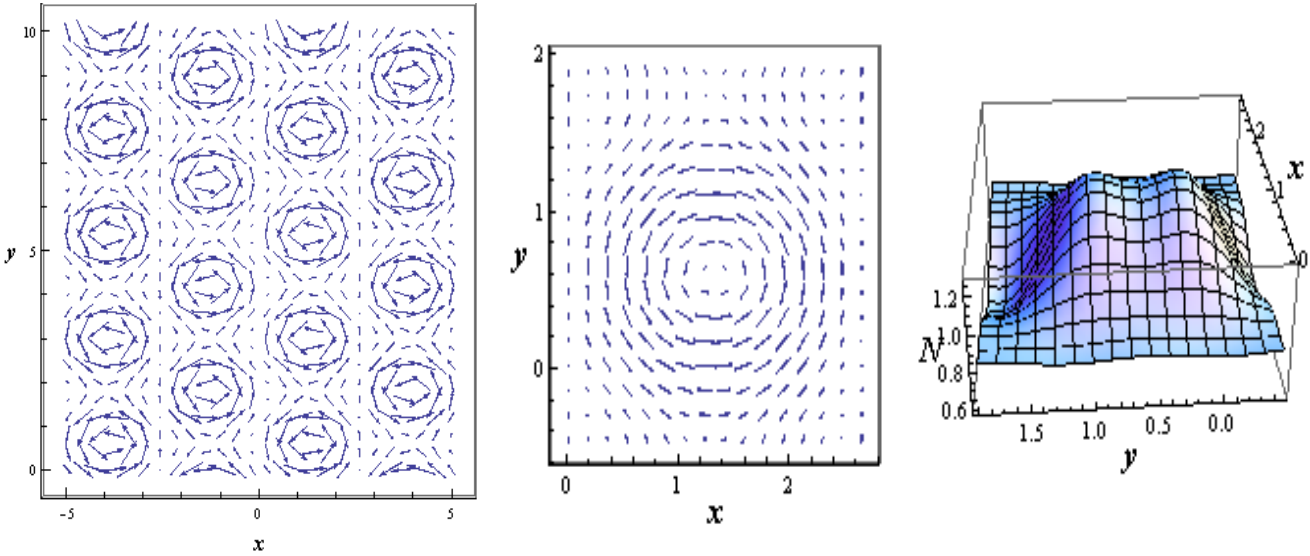

**Fig. 5** Plot of vortex structure (left), single skyrmion (center) and absolute normalization (right) $N$ for the first solution based on the corrected solution (9).

We sum up our results. We obtained the lowest energy approximate solutions of the complete skymion field equations and their corresponding vortex structures. Two solutions emerged with the vortex swirls in opposite directions leading to an energy difference of 2 meV per skyrmion pair. Comparing this with the much larger gap $\Delta \approx 1$ eV usually assumed for ferromagnets [20] could account for the extremely low current densities observed in current-induced experiments with skyrmions. The solutions are associated with an extremum property, which favors an array of almost hexagonal shape. We predict that a *regular* hexagonal lattice must have a mix of skyrmions of both swirls.

Figure 1 suggests other interesting scenarios (e.g., the zeros of $G$) for further study. Although the norm of the spins could not be kept constant at unity, we found that their greatest deviation from unity occurred in regions where the spins are not planar. The correction to the solution $\Psi_0$ is precisely due to the Dzyaloshinkii-Moriya (DM) coupling. We saw from Fig. 5 that this correction improved the vortex solution's normalization. We had studied only the $n = 0$ solutions so it would be interesting to consider higher $n$ values.